\documentclass[pra,aps,twocolumn,superscriptaddress,showpacs,floatfix]{revtex4}
\usepackage{amsmath}
\usepackage{graphicx}
\usepackage{simon}

\begin{document}

\title{Nonlinear magneto-optical rotation of frequency-modulated light resonant with a low-$J$ transition}
\author{Yu.\ P. Malakyan}\email{yumal@ipr.sci.am}
\affiliation{Institute for Physical Research, National Academy of
Sciences of Armenia, Ashtarak-2, 378410, Armenia}
\author{S. M. Rochester}\email{simonkeys@yahoo.com}
\affiliation{Department of Physics, University of California at
Berkeley, Berkeley, California 94720-7300}
\author{D. Budker}\email{budker@socrates.berkeley.edu}
\affiliation{Department of Physics, University of California at
Berkeley, Berkeley, California 94720-7300} \affiliation{Nuclear
Science Division, Lawrence Berkeley National Laboratory, Berkeley,
California 94720}
\author{D. F. Kimball}\email{dfk@uclink4.berkeley.edu}
\affiliation{Department of
Physics, University of California at Berkeley, Berkeley,
California 94720-7300}
\author{V. V. Yashchuk}\email{yashchuk@socrates.berkeley.edu}
\affiliation{Department of Physics, University of California at
Berkeley, Berkeley, California 94720-7300}

\date{\today}

\begin{abstract}
A low-light-power theory of nonlinear magneto-optical rotation of
frequency-modulated light resonant with a $J=1\ra J'=0$ transition
is presented. The theory is developed for a Doppler-free
transition, and then modified to account for Doppler broadening
and velocity mixing due to collisions. The results of the theory
are shown to be in qualitative agreement with experimental data
obtained for the rubidium $D1$ line.
\end{abstract}



\maketitle

\section{Introduction}
\label{Sec:Intro}

\begin{figure}
    \includegraphics{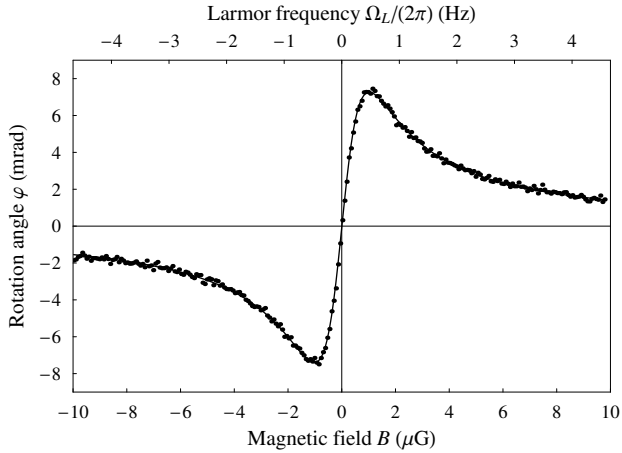}
    \caption{Experimental (dots) and dispersive-Lorentzian
   least-squares fit (line) magnetic-field dependence of NMOR in a
   10-cm-diameter paraffin-coated $^{85}$Rb-vapor cell, obtained
   as described in Ref. \cite{Bud2000Sens}. Linearly polarized
   laser light is tuned to the high-frequency side of the $D2$
   $F=3 \rightarrow F'$ transition, at which maximum rotation
   occurs. The light intensity is $\sim$50 ${\rm \Gm W\,cm^{-2}}$,
   and the beam diameter is $\sim$2 mm. The temperature of the
   cell is $\sim$$19^\circ$C corresponding to a vapor density of
   $\sim$$4\times10^9{\rm cm^{-3}}$.}
    \label{Fig:ZeroFieldData}
\end{figure}

Nonlinear magneto-optical rotation (NMOR), or
light-power-dependent rotation of optical polarization due to
resonant interaction with an atomic medium in the presence of a
magnetic field $B$, has applications ranging from fundamental
symmetry tests to magnetometry \cite{Bud2002RMP}. With NMOR due to
the evolution of ground-state atomic polarization \cite{Bud99},
optical rotation is proportional to the magnetic field for small
fields, but falls off when the Larmor frequency $\GO_L=g\Gm_0B$
($g$ is the gyromagnetic ratio, $\Gm_0$ is the Bohr magneton, and
we set $\hbar=1$ throughout) becomes larger than half of the
atomic polarization relaxation rate $\Gg$ (Fig.\
\ref{Fig:ZeroFieldData}).
Atomic polarization relaxation rates as low as
$\Gg\simeq2\Gp\!\times1$ Hz can be achieved for alkali atoms
contained in paraffin-coated vapor cells \cite{AlexandrovLPh96},
corresponding to magnetic field widths of approximately 1 $\Gm$G
\cite{Bud98} and high magnetometric sensitivity ($\sim$3
$\rm{pG\,Hz^{-1/2}}$ \cite{Bud2000Sens}) to small fields.

\begin{figure}
    \includegraphics{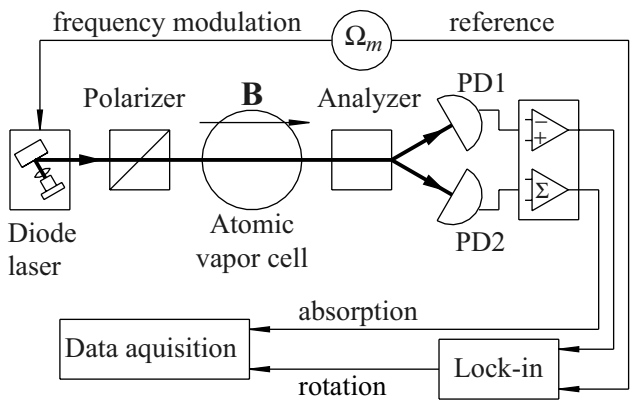}
    \caption{Simplified schematic of the apparatus used to detect
    FM NMOR signals. A paraffin-coated cell containing Rb vapor is
    placed inside a balanced polarimeter (a polarizer and an
    analyzer oriented at $\sim$$45^\circ$ with respect to each
    other). The frequency of the laser is modulated with an
    amplitude of a few dozen MHz. The lock-in
    amplifier is used to detect the components of optical
    rotation oscillating both in phase and $\Gp/2$ out of phase with
    the frequency modulation.} \label{Fig:FMNMORsetup}
\end{figure}

With a traditional NMOR magnetometer, the high small-field
sensitivity comes at the expense of a limited dynamic range. Since
many applications (such as measurement of geomagnetic fields or
magnetic fields in space \cite{Rip2001}) require high sensitivity
at magnetic fields on the order of a Gauss, a method to extend the
magnetometer's dynamic range is needed. It was recently
demonstrated \cite{Bud2002FM,Yas2003Select} that when
frequency-modulated light is used to induce and detect nonlinear
magneto-optical rotation (FM NMOR), the narrow features in the
magnetic-field dependence of optical rotation normally centered at
$B=0$ can be translated to much larger magnetic fields. In this
setup (Fig.\ \ref{Fig:FMNMORsetup}), the light frequency is
modulated at frequency $\GO_m$, and the time-dependent optical
rotation is measured at a harmonic of this frequency.
Narrow features appear, centered at Larmor frequencies that are
integer multiples of $\GO_m$, allowing the dynamic range of the
magnetometer to extend well beyond the Earth field.

Light-frequency modulation has been previously applied to
measurements of \emph{linear} magneto-optical rotation and
parity-violating optical rotation \cite{Bar78,Bark88b} in order to
produce a time-dependent optical rotation signal without
introducing additional optical elements (such as a Faraday
modulator) between the polarizer and analyzer. Optical pumping
with frequency-modulated light has been applied to magnetometry
with $^4$He \cite{Che95,Che96,Gil2001} and Cs \cite{And2003}; in
these experiments transmission, rather than optical rotation, was
monitored. In the latter work with Cs, the modulation index (the
ratio of modulation depth to modulation frequency) is on the order
of unity, in contrast to the much larger index in the work
described here, allowing interpretation of the process in terms of
the $\GL$- or coherent-population-trapping resonances. This regime
has also been explored in Rb using modulation of the magnetic
field, rather than the light field \cite{Val2003}. The closely
related method of modulation of light intensity (synchronous
optical pumping) predates the frequency-modulation technique
\cite{Bel61a}. Also employing light-intensity modulation is the
so-called quantum beat resonance technique \cite{Ale63} used, for
example, for measuring the Land\'{e} factors of molecular ground
states (see Ref.\ \cite{Auz95} and references therein). Intensity
modulation was recently used in experiments that put an upper
limit on the (parity- and time-reversal-violating) electric dipole
moment of $^{199}$Hg (Refs.\ \cite{Rom2001PRL,Rom2001} and
references therein).

A quantitative theory of FM NMOR would be of use in the study and
application of the technique. As a first step towards a complete
theory, we present here a perturbative calculation for a $J=1 \ra
J'=0$ atomic transition that takes into account Doppler broadening
and averaging due to velocity-changing collisions. We begin the
discussion in Sec.\ \ref{Sec:DataAndTheory} by comparing
experimental FM NMOR magnetic-field-dependence data obtained with
a paraffin-coated $^{87}$Rb-vapor cell to the predictions of the
calculation (described in Sec.\ \ref{Sec:Theory}). We find that
the simplified model still reproduces the salient features of the
observed signals, indicating that the magnetic-field dependence of
FM NMOR at low light power is not strongly dependent on power or
angular momentum. As discussed in Sec.\ \ref{Sec:Conclusion}, the
description of the saturation behavior and spectrum of FM NMOR in
a system like Rb, on the other hand, will require a more complete
theory.

\section{Experimental data and comparison with theory}
\label{Sec:DataAndTheory}

\begin{figure}
    \includegraphics{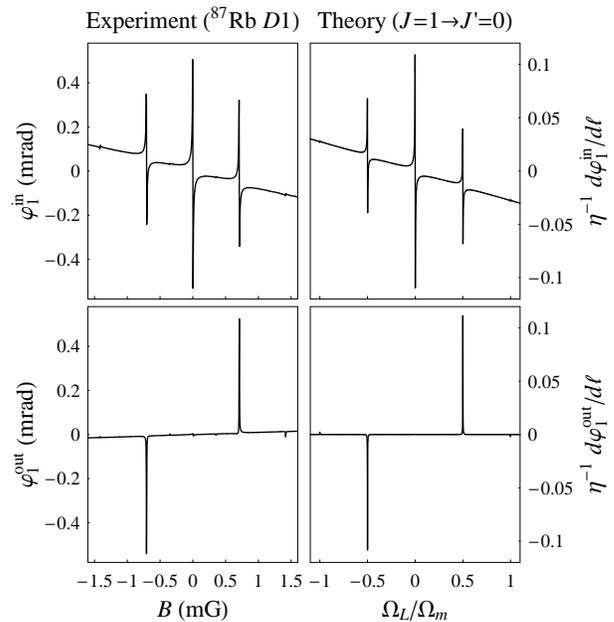}
    \caption{Measured (left column) and calculated (right column)
    in-phase (top row) and quadrature (bottom row) first-harmonic
    amplitudes of FM NMOR. The experimental signals, plotted as a
    function of magnetic field $B$ applied along the light
    propagation direction, are obtained with light tuned to the
    wing of the $F=2\ra F'=1$ absorption line of the $^{87}$Rb
    $D1$ spectrum. The laser power is 15 $\Gm$W, beam diameter is
    $\sim$2 mm, $\GO_m=2\Gp\times1$ kHz, and modulation amplitude
    is $2\pi\times220$ MHz. All resonances have widths ($\sim$1
    $\Gm$G) corresponding to the rate of atomic polarization
    relaxation in the paraffin-coated cell. The normalized
    calculated signals [Eq.\ \ref{Eq:sigTot}], for a $J=1\ra J'=0$
    transition, are plotted as a function of normalized Larmor
    frequency $\GO_L/\GO_m$. For these plots, the parameters
    $\GD_0/\GG_{\!D}=0.7$, $\GO_m/\Gg=500$, and $\GD_l/\GD_0=1$
    (described in Sec.\ \ref{Sec:Theory}) are chosen to match the
    experimental parameters given above.}
    \label{Fig:1HDataTheoryPlot}
\end{figure}

\begin{figure}
    \includegraphics{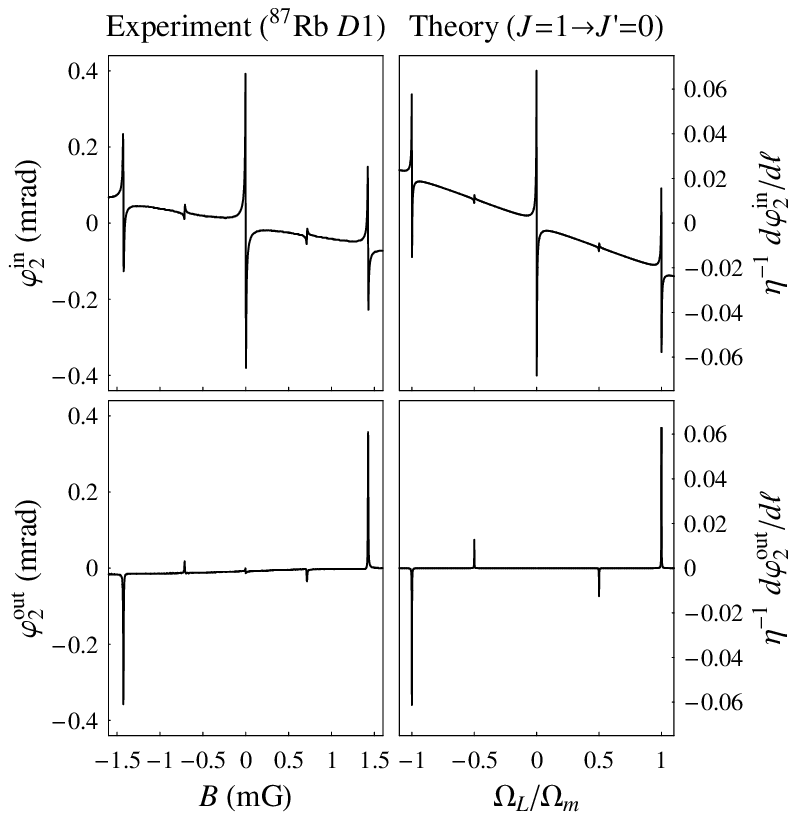}
    \caption{Measured and calculated second-harmonic amplitudes of
    FM NMOR. See caption to Fig. \ref{Fig:1HDataTheoryPlot}. For
    the experimental signals, light is tuned to the center of the
    $F=2\ra F'=1$ absorption line of the $^{87}$Rb $D1$ spectrum,
    the laser power is 15 $\Gm$W, beam diameter is $\sim$2.5 mm,
    $\GO_m=2\Gp\times1$ kHz, and modulation amplitude is
    $2\pi\times440$ MHz. The parameters for the theoretical
    signals are $\GD_0/\GG_{\!D}=1.4$, $\GO_m/\Gg=500$, and
    $\GD_l/\GD_0=0.2$.} \label{Fig:2HDataTheoryPlot}
\end{figure}

Figures \ref{Fig:1HDataTheoryPlot} and \ref{Fig:2HDataTheoryPlot}
show first- and second-harmonic data, respectively, obtained from
an FM NMOR magnetometer with light tuned near the $D1$ line of
rubidium in the manner described in Ref.\
\cite{Bud2002FM,Yas2003Select}, along with the predicted signals
for a $J=1\ra J'=0$ transition obtained from the theory described
in Sec.\ \ref{Sec:Theory} with parameters matching those of the
experimental data.
The calculation for the simpler system reproduces many of the
qualitative aspects of the experimental data for Rb. The features
at the center of the in-phase plots of Figs.\
\ref{Fig:1HDataTheoryPlot} and \ref{Fig:2HDataTheoryPlot} are the
zero-field resonances, analogous to the one shown in Fig.\
\ref{Fig:ZeroFieldData}. (The background linear slope seen in the
in-phase signals is also a zero-field resonance, due to the
``transit effect'' \cite{Bud2002RMP}. It is modelled in the theory
by an extra term analogous to the others with the isotropic
relaxation rate $\Gg$ equal to the transit rate of atoms through
the laser beam.) In addition to these features, there appear new
features centered at magnetic field values at which
$\abs{\GO_L/\GO_m}=1/2$ and 1. For the first-harmonic signal, the
former are larger, whereas for the second-harmonic, the latter
are; this is primarily a result of the different light detunings
used in the two measurements. For these new resonances, there are
both dispersively shaped in-phase signals and $\pi/2$ out of phase
(quadrature) components peaked at the centers of these resonances.
The resonances occur when the optical pumping rate, which is
periodic with frequency $\GO_m$ due to the laser frequency
modulation, is synchronized with Larmor precession, which for an
aligned state has periodicity at frequency $2\GO_L$ as a result of
the state's rotational symmetry. This results in the atomic medium
being optically pumped into an aligned rotating state, modulating
the optical properties of the medium at $2\GO_L$. The aligned
atoms produce maximum optical rotation when the alignment axis is
at $\pi/4$ to the direction of the light polarization and no
rotation when the axis is along the light polarization. Thus, on
resonance, there is no in-phase signal and maximum quadrature
signal. The relative sizes and signs of the features in the
magnetic-field dependence, largely determined by the ratio of the
modulation width $\GD_0$ to the Doppler width $\GG_{\!D}$ (Sec.\
\ref{Sec:Theory}), are well reproduced by the theory. The theory
also exhibits the expected linear light-power dependence of the
optical rotation amplitude as observed in experiments at low power
\cite{Bud2002FM,Yas2003Select}.

There are additional features, centered at
$\abs{\GO_L/\GO_m}=1/4$, just barely visible in the experimental
plots of Figs.\ \ref{Fig:1HDataTheoryPlot},
\ref{Fig:2HDataTheoryPlot}. These features, which become more
prominent at higher light power \cite{Yas2003Select}, are due to
the optical pumping, precession, and detection of the hexadecapole
moment. These resonances are not described by the current theory,
because the presence of the hexadecapole moment requires
ground-state angular momentum $J\ge2$ and second-order light
interactions. A quantitative description of these resonances is
among the goals for an expanded theory.

\section{Theory}
\label{Sec:Theory}

\subsection{Introduction}

\begin{figure}
    \includegraphics{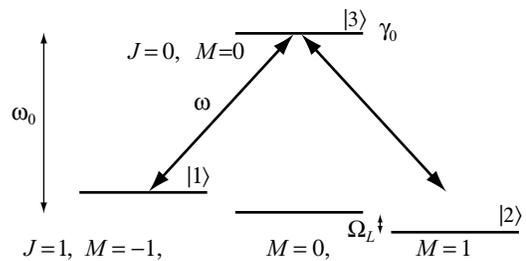}
    \caption{A $J_g=1\ra J_e=0$ atomic transition of frequency
    $\Go_0$. The lower sublevels are split by the Larmor frequency
    $\GO_L$. The arrows indicate the interaction with light of
    frequency $\Go$ polarized perpendicular to the quantization
    axis. The upper state spontaneously decays a rate $\Gg_0$.}
    \label{Fig:LevelDiagram}
\end{figure}

The goals for a complete theory of FM NMOR are outlined in Sec.\
\ref{Sec:Conclusion}. As a first step towards such a theory, we
calculate here the optical rotation due to interaction of
frequency-modulated light with a $J_g=1\ra J_e=0$ atomic
transition (Fig.\ \ref{Fig:LevelDiagram}), where the subscripts
$g$ and $e$ indicate the ground and excited states, respectively.
We will assume that the light power is low enough that no optical
pumping saturation occurs.

We begin by calculating the time-dependent atomic ground-state
coherence of a Doppler-free system. Using the magnetic-field--atom
and light--atom interaction Hamiltonians (under the rotating wave
approximation) we write the density-matrix evolution equations.
Under the low-light-power approximation, an expression for the
ground-state atomic coherence can be written as a time integral.
We convert the integral to a sum over harmonics of the modulation
frequency by expanding the integrand as a series. This form is
convenient for this calculation because the optical rotation
signal is measured by lock-in detection. The expression for the
Doppler-free case is then modified to take into account Doppler
broadening and velocity averaging due to collisions. Atoms in an
antirelaxation-coated vapor cell collide with the cell walls in
between interactions with the light beam, preserving their
polarization but randomizing their velocities. In the
low-light-power case, we can account for this by first calculating
the effect of optical pumping assuming no collisions, and then
averaging the density matrix over atomic velocity. Note that in
this case we assume that optical pumping is unsaturated not only
for the resonant velocity group, but also when atomic polarization
is averaged over the velocity distribution and cell volume.

Using the wave equation, we find an expression for the
time-dependent optical rotation in terms of the atomic
ground-state coherence of a given atomic velocity group. This
rotation is then integrated over time and atomic velocity to
obtain an expression for the signal at a given harmonic measured
by the lock-in detector.

\subsection{The Hamiltonian}

The total Hamiltonian $H$ is the sum of the unperturbed
Hamiltonian $H_0$, the light--atom-interaction Hamiltonian $H_l$,
and the magnetic-field--atom-interaction Hamiltonian $H_B$. Using
the basis states $\ket{\xi J M}$, where $\xi$ represents
additional quantum numbers, denoted by
\begin{equation}
\begin{alignedat}{3}
    \ket{\xi_gJ_g,-1}&=\begin{pmatrix}1\\0\\0\\0\end{pmatrix},&\quad
    \ket{\xi_gJ_g,0}&=\begin{pmatrix}0\\1\\0\\0\end{pmatrix},\\
    \ket{\xi_gJ_g,1}&=\begin{pmatrix}0\\0\\1\\0\end{pmatrix},&\quad
    \ket{\xi_eJ_e,0}&=\begin{pmatrix}0\\0\\0\\1\end{pmatrix},
\end{alignedat}
\end{equation}
the unperturbed Hamiltonian $H_0$ is given by
\begin{equation}
    H_0=\begin{pmatrix}
            0&0&0&0\\
            0&0&0&0\\
            0&0&0&0\\
            0&0&0&\Go_0
        \end{pmatrix},
\end{equation}
where $\Go_0$ is the transition frequency (again, we set $\hbar=1$
throughout).

An $x$-polarized optical electric field $\mb{E}$ is written as
\begin{equation}
    \mb{E}=E_0\cos\prn{\Go t}\hat{\mb{e}}_x.
\end{equation}
where $E_0$ is the electric field amplitude and $\Go$ is the
frequency (modulated as $\Go=\Go_l-\GD_0\cos\GO_mt$, where $\Go_l$
is the laser carrier frequency and $\GD_0$ is the modulation
amplitude). We assume that the atomic medium is optically thin, so
that we can neglect the change in light polarization and intensity
inside the medium when calculating the state of the medium. The
light-atom interaction Hamiltonian is given by
\begin{equation}
\begin{split}
    H_l &{}=-\mb{E}\cdot\mb{d}\\
        &{}=-E_0\cos\prn{\Go t}d_x\\
        &{}=-\frac{1}{\sqrt{2}}E_0\cos\prn{\Go
        t}\prn{d_{-1}-d_{+1}},
\end{split}
\end{equation}
where $\mb{d}$ is the dipole operator. According to the
Wigner-Eckart theorem, components of a tensor operator $T_{\Gk q}$
are related to the reduced matrix element $\rme{\xi J}{T_\Gk}{\xi
J'}$ by \cite{Sob92}
\begin{equation}
    \bra{\xi\!JM}T_{\Gk q}\ket{\xi'\ms{-4}J'\!M'}
        =(-1)^{J-M}\!\rme{\xi\!J}{T_\Gk}{\xi'\ms{-4}J'}\!
            \threej(J,-M)(\Gk,q)(J',M')\ms{-5}.
\end{equation}
Thus the matrix elements of $d_{+1}$ and $d_{-1}$ for this
transition can be written
\begin{equation}
\begin{split}
    &\bra{\xi\!JM}d_{\pm1}\ket{\xi'\ms{-4}J'\!M'}
        =(-1)^{J-M}\!\rme{\xi\!J}{d}{\xi'\ms{-4}J'}\!
            \threej(J,-M)(1,\pm1)(J',M')\\
        &=\frac{1}{\sqrt{3}}\times
            \begin{cases}
                \rme{\xi_eJ_e}{d}{\xi_gJ_g} &\text{for $\xi=\xi_e,J=J_e,M=0$}\\
                                        &\text{and $\xi'=\xi_g,J'=J_g,M'=\mp1$,}\\
                \rme{\xi_gJ_g}{d}{\xi_eJ_e} &\text{for $\xi=\xi_g,J=J_g,M=\pm1$}\\
                                        &\text{and $\xi'=\xi_e,J'=J_e,M'=0$,}\\
                0&\text{in all other cases.}
            \end{cases}
\end{split}
\end{equation}
Reduced matrix elements with different ordering of states are
related by \cite{Sob92}
\begin{equation}
    \rme{\xi\!J}{T_\Gk}{\xi'\ms{-4}J'}=(-1)^{J-J'}\rme{\xi'\ms{-4}J'}{T_\Gk}{\xi\!J}^*,
\end{equation}
and since the reduced dipole matrix element is real,
\begin{equation}
    \rme{\xi_gJ_g}{d}{\xi_eJ_e}=-\rme{\xi_eJ_e}{d}{\xi_gJ_g}.
\end{equation}
Thus $H_l$ is given in matrix form by
\begin{equation}
    H_l=2\GO\cos\Go t
        \begin{pmatrix}
            0&0&0&-1\\
            0&0&0&0\\
            0&0&0&1\\
            -1&0&1&0
        \end{pmatrix},
\end{equation}
where $\GO=\rme{\xi_gJ_g}{d}{\xi_eJ_e}E_0/\!\prn{2\sqrt{6}\,}$ is
(apart from a numerical factor of order unity) the optical Rabi
frequency.

The magnetic field interaction Hamiltonian $H_B$ for a
$\hat{\mb{z}}$-directed magnetic field $\mb{B}$ is given by
\begin{equation}
\begin{split}
    H_B
    &{}=-\bs{\Gm}\cdot\mb{B}\\
    &{}=g\Gm_0\,\mb{J}\cdot\mb{B}\\
    &{}=g\Gm_0 J_z B\\
    &{}=\GO_L
        \begin{pmatrix}
            -1&0&0&0\\
            0&0&0&0\\
            0&0&1&0\\
            0&0&0&0
        \end{pmatrix},
\end{split}
\end{equation}
where $\GO_L$ is the Larmor frequency as defined in Sec.\
\ref{Sec:Intro}. Thus, the total Hamiltonian is given by
\begin{equation}
\begin{split}
    H
    &{}=H_0+H_l+H_B\\
    &{}=
        \begin{pmatrix}
            -\GO_L&0&0&-2\GO\cos\Go t\\
            0&0&0&0\\
            0&0&\GO_L&2\GO\cos\Go t\\
            -2\GO\cos\Go t&0&2\GO\cos\Go t&\Go_0
        \end{pmatrix}.
\end{split}
\end{equation}

\subsection{Rotating-wave approximation}

We now use the rotating-wave approximation in order to remove the
optical-frequency time dependence from the Hamiltonian. We first
transform into the frame rotating at the optical frequency by
means of the unitary transformation operator
$U(t)=\exp\prn{-iH't}$, where
\begin{equation}
    H'=
        \begin{pmatrix}
            0  &0  &0   &0\\
            0  &0  &0   &0\\
            0  &0  &0   &0\\
            0  &0  &0   &\Go
        \end{pmatrix}
\end{equation}
is the unperturbed Hamiltonian $H_0$ with $\Go_0$ replaced by
$\Go$. It is straightforward to show that under this
transformation the Hamiltonian in the rotating frame is given by
\begin{equation}
\begin{split}
    \tilde{H}
        &{}=U^{-1}(t)\,H(t)\,U(t)-i\,U^{-1}(t)\,\frac{d}{dt}\,U(t)\\
        &{}=
        \begin{pmatrix}
            -\GO_L          &0  &0              &-\GO\prn{1+e^{-2i\Go t}}\\
            0               &0  &0              &0\\
            0               &0  &\GO_L          &\GO\prn{1+e^{-2i\Go t}}\\
            -\GO\prn{1+e^{2i\Go t}}&0  &\GO\prn{1+e^{2i\Go{}t}}&\Go_0-\Go
        \end{pmatrix},
\end{split}
\end{equation}
where we have used $\cos\Go t=\prn{e^{-i\Go t}+e^{i\Go t}}\!/\,2$.
Averaging over an optical cycle to remove far-off-resonant terms
(the rotating wave approximation), we have
\begin{equation}
    \tilde{H}\simeq
        \begin{pmatrix}
            -\GO_L          &0  &0              &-\GO\\
            0               &0  &0              &0\\
            0               &0  &\GO_L          &\GO\\
            -\GO            &0  &\GO            &-\GD
        \end{pmatrix}.
\end{equation}
where $\GD=\Go-\Go_0$ is the (time-dependent) optical detuning.

\subsection{Relaxation and repopulation}

We assume that the upper state spontaneously decays with a rate
$\Gg_0$, and that the ground state relaxes with a rate $\Gg$, due
to the exit of atoms from the light beam, in the case of the
``transit'' effect, or collisions with other atoms or the cell
wall in the case of the ``wall-induced Ramsey effect''
\cite{Bud2002RMP}. (Additional upper-state relaxation processes
can be neglected in comparison with the spontaneous decay rate.)
This relaxation is described by the matrix $\GG$, given by
\begin{equation}
    \GG=
        \begin{pmatrix}
            \Gg     &0      &0      &0\\
            0       &\Gg    &0      &0\\
            0       &0      &\Gg    &0\\
            0       &0      &0      &\Gg_0
        \end{pmatrix}.
\end{equation}
The simplest model of ground state relaxation is used. The effects
of collisional dephasing could be included by adding off-diagonal
terms if a more realistic model is desired. In order to conserve
the number of atoms, the ground state must be replenished at the
same rate at which it relaxes. This is described by the
repopulation matrix $\GL$:
\begin{equation}
    \GL=\frac{N}{3}
        \begin{pmatrix}
            \Gg     &0      &0      &0\\
            0       &\Gg    &0      &0\\
            0       &0      &\Gg    &0\\
            0       &0      &0      &0
        \end{pmatrix},
\end{equation}
where $N$ is the atomic density. We ignore repopulation due to
spontaneous decay since the calculation is performed in the
low-light-power limit ($\GO^2\ll\Gg_0\Gg$).

\subsection{Density-matrix evolution equations}

The evolution of the density matrix $\Gr$ (defined so that
$\tr\Gr=N$) is given by the Liouville equation \cite{Scu97}
\begin{equation}\label{Eq:Liou}
    \dot{\Gr}=-i\sbr{\tilde{H},\Gr}-\frac{1}{2}\cbr{\GG,\Gr}+\GL,
\end{equation}
where the square brackets denote the commutator and the curly
brackets the anticommutator. The $M=0$ ground-state sublevel does
not couple to the light, and can be ignored. Using $\ket{1}$ and
$\ket{2}$ to denote the ground-state $M=-1$ and $+1$ sublevels,
respectively, and $\ket{3}$ to denote the upper state, and
assuming that $\Gg\ll\Gg_0$, the evolution equations for the
atomic coherences obtained from Eq.\ (\ref{Eq:Liou}) are
\begin{subequations}
\begin{gather}
    \dot{\Gr}_{31}=-\sbr{\Gg_0/2+i\prn{\GO_L-\GD}}\Gr_{31}+i\GO\prn{\Gr_{11}-\Gr_{33}-\Gr_{21}},\label{EvolEq31}\\
    \dot{\Gr}_{23}=-\sbr{\Gg_0/2+i\prn{\GO_L+\GD}}\Gr_{23}+i\GO\prn{\Gr_{22}-\Gr_{33}-\Gr_{21}},\label{EvolEq32}\\
    \dot{\Gr}_{21}=-\prn{\Gg+2i\GO_L}\Gr_{21}-i\GO\prn{\Gr_{31}+\Gr_{23}}.\label{EvolEq21}
\end{gather}
\end{subequations}
We can assume that in the low-light-power limit the populations
$\Gr_{11,22,33}$ are essentially unperturbed by the light
($\Gr_{11,22}\simeq N/3$, $\Gr_{33}\ll N$). We can also assume
that, neglecting transient terms, the optical coherences
$\Gr_{31,23}$ are slowly varying (any time dependence would be due
to modulation of the light frequency, which will always be done at
a rate much less than $\Gg_0$; thus
$\dot{\Gr}_{31,23}\ll\Gg_0\Gr_{31,23}$). Using these assumptions,
the evolution equations for the atomic coherences [Eqs.\
(\ref{EvolEq31}--\ref{EvolEq21})] become
\begin{subequations}
\begin{gather}
    0\simeq-\sbr{\Gg_0/2+i\prn{\GO_L-\GD}}\Gr_{31}+i\GO\prn{N/3-\Gr_{21}},\label{EvolEq31a}\\
    0\simeq-\sbr{\Gg_0/2+i\prn{\GO_L+\GD}}\Gr_{23}+i\GO\prn{N/3-\Gr_{21}},\label{EvolEq32a}\\
    \dot{\Gr}_{21}{}\simeq-\prn{\Gg+2i\GO_L}\Gr_{21}-i\GO\prn{\Gr_{31}+\Gr_{23}}.\label{EvolEq21a}
\end{gather}
\end{subequations}
These equations can be used to solve for the optical and
ground-state coherences.

\subsection{Calculation of the optical and ground-state coherences}

The expression for optical rotation (Sec.\ \ref{Sec:Signal}) is
written in terms of the optical coherences $\Gr_{31,23}$. We will
now relate the optical coherences to the ground-state coherence
$\Gr_{21}$ and find an expression for $\Gr_{21}$ as a sum over
harmonics of the light detuning modulation frequency $\GO_m$. This
form is convenient because the signal is measured at harmonics of
this frequency.

Solving Eqs.\ (\ref{EvolEq31a}) and (\ref{EvolEq32a}) for
$\Gr_{31}$ and $\Gr_{23}$ in terms of $\Gr_{21}$, we obtain
\begin{equation}\label{OptCoherence}
\begin{split}
    \Gr_{31}&{}\simeq\frac{\GO\prn{N/3-\Gr_{21}}}{\GO_L-\GD-i\Gg_0/2}\,,\\
    \Gr_{23}&{}\simeq\frac{\GO\prn{N/3-\Gr_{21}}}{\GO_L+\GD-i\Gg_0/2}\,.
\end{split}
\end{equation}
In order to solve for $\Gr_{21}$, we make the substitution
$\Gr_{21}\ra r_{21} e^{-\prn{2i\GO_L+\Gg}t}$ in Eq.\
(\ref{EvolEq21a}):
\begin{equation}
    \dot{r}_{21}\simeq-i\GO\prn{\Gr_{31}+\Gr_{23}}e^{\prn{2i\GO_L+\Gg}t},
\end{equation}
or, integrating (assuming that $r_{21}=\Gr_{21}=0$ at $t=0$),
\begin{equation}
    r_{21}\simeq-i\GO
    \int_0^t\prn{\Gr_{31}+\Gr_{23}}e^{\prn{2i\GO_L+\Gg}\Gt}d\Gt,
\end{equation}
so, substituting back,
\begin{equation}\label{GroundCoherence}
    \Gr_{21}\simeq-i\GO
    \int_0^t
    \prn{\Gr_{31}+\Gr_{23}}e^{-\prn{2i\GO_L+\Gg}\prn{t-\Gt}}d\Gt.
\end{equation}
The expressions for the optical coherences [Eqs.\
(\ref{OptCoherence})] are then substituted into the expression for
the ground-state coherence [Eq.\ (\ref{GroundCoherence})].
Assuming that the light power is low ($\GO\ll\Gg_0$) allows us to
neglect second-order terms. We also assume that the level shift
induced by the magnetic field is smaller than the natural line
width, i.e. $\GO_L\ll\Gg_0$. (For the $D$-lines of rubidium used
in the experiment, this assumption holds for magnetic fields up to
the earth-field range.) The ground-state coherence is then given
by
\begin{equation}\label{r21Eq}
\begin{split}
    \Gr_{21}
    &{}\simeq-\frac{i}{3}\,\GO^2N\!
        \int_0^t
            \prn{\frac{1}{\GO_L-\GD-i\Gg_0/2}+\frac{1}{\GO_L+\GD-i\Gg_0/2}}\\
            &\ms{250}\times e^{-\prn{2i\GO_L+\Gg}\prn{t-\Gt}}
        d\Gt\\
    &{}\simeq\frac{2}{3}\,\GO^2N\!
        \int_0^t
            \prn{
                \frac{\Gg_0/2-i\GO_L}{\GD\ms{-1}^2+\Gg_0^2/4}
                +\frac{2i\GO_L\GD\ms{-1}^2}{\prn{\GD\ms{-1}^2+\Gg_0^2/4}^2}
            }\\
            &\ms{250}\times e^{-\prn{2i\GO_L+\Gg}\prn{t-\Gt}}
        d\Gt\\
    &=\frac{2}{3}\,\GO^2N\sbr{\prn{\Gg_0/2-i\GO_L}I_1(t)+2i\GO_L I_2(t)},
\end{split}
\end{equation}
where the integral $I_1$ has been defined by
\begin{equation}\label{Eq:I1}
\begin{split}
    I_1(t)
        &=\int_0^t
            \frac{e^{-\prn{2i\GO_L+\Gg}\prn{t-\Gt}}\,d\Gt}{\GD_0^2\prn{D_0-\cos\GO_m\Gt}^2+\Gg_0^2/4}\,
            \\
        &=\prn{\Gg_0^2/4}^{-1}
            \int_0^t
                f\!\prn{\GO_m\Gt}\,e^{-\prn{2i\GO_L+\Gg}\prn{t-\Gt}}\,d\Gt,
\end{split}
\end{equation}
and $I_2$ has been defined by
\begin{equation}\label{Eq:I2}
\begin{split}
    I_2(t)
        &=\ms{-5}\int_0^t\ms{-7}
            \frac{\GD_0^2\prn{D_0-\cos\GO_m\Gt}^2d\Gt}{\sbr{\GD_0^2\prn{D_0-\cos\GO_m\Gt}^2+\Gg_0^2/4}^2}\,
            e^{-\prn{2i\GO_L+\Gg}\prn{t-\Gt}}\\
        &=-\GD_0^2\frac{\partial I_1(t)}{\partial\GD_0^2}.
\end{split}
\end{equation}
Here we have substituted for $\GD$ the expression for the
light-frequency modulation $\GD=\GD_0\prn{D_0-\cos\GO_m\Gt}$,
where the dimensionless average detuning parameter $D_0$ is
defined by $D_0=\GD_l/\GD_0$, where $\GD_l=\Go_l-\Go_0$. The
lineshape factor $f(x)$ is defined by
\begin{equation}\label{Eq:fx}
    f(x)=\frac{\Gg_0^2/4}{\GD_0^2\prn{D_0-\cos x}^2+\Gg_0^2/4}.
\end{equation}
Expanding this function as a series of harmonics,
\begin{equation}\label{Eq:seriesExp}
    f(x)=\sum_{n=-\infty}^{\infty}a_n e^{inx},
\end{equation}
the coefficients $a_n$ are given by
\begin{equation}\label{Eq:anExact}
    a_n
    =\frac{1}{2\Gp}
        \int_{-\Gp}^{\Gp}f(x)\,\cos nx\,dx.
\end{equation}
Substituting the series expansion for $f(x)$ into $I_1$, we have
\begin{equation}\label{Eq:I1Expr}
\begin{split}
    I_1(t)
    &{}\simeq
        \int_0^t
            \sbr{
                \prn{\Gg_0^2/4}^{-1}\ms{-8}
                \sum_{n=-\infty}^{\infty}a_n e^{in\GO_m\Gt}
            }
            e^{-\prn{2i\GO_L+\Gg}\prn{t-\Gt}}
        d\Gt\\
    &=\prn{\Gg_0^2/4}^{-1}\ms{-8}
        \sum_{n=-\infty}^{\infty}
            \frac{e^{in\GO_mt}-e^{-\prn{\Gg+2i\GO_L}t}}{\Gg+i\prn{2\GO_L+n\GO_m}}\,a_n\\
    &{}\simeq
        \prn{\Gg_0^2/4}^{-1}\ms{-8}
        \sum_{n=-\infty}^{\infty}
            \frac{a_ne^{in\GO_mt}}{\Gg+i\prn{2\GO_L+n\GO_m}}\,,
\end{split}
\end{equation}
where we have discarded the transient term
$e^{-\prn{\Gg+2i\GO_L}t}$. The expression for $I_2$ [Eq.\
(\ref{Eq:I2})] can be found from that for $I_1$:
\begin{equation}
\begin{split}
    I_2(t)
    &=-\GD_0^2\frac{\partial I_1(t)}{\partial\GD_0^2}\\
    &{}\simeq
        \prn{\Gg_0^2/4}^{-1}\ms{-8}
        \sum_{n=-\infty}^{\infty}
            \frac{b_ne^{in\GO_mt}}{\Gg+i\prn{2\GO_L+n\GO_m}}\,,
\end{split}
\end{equation}
where the coefficient $b_n$ is defined by
\begin{equation}\label{Eq:bnDef}
    b_n
    =-\GD_0^2\frac{\partial a_n}{\partial\GD_0^2}
\end{equation}
In order to find the relative values of $a_n$ and $b_n$, it is
useful to have an approximate expression for them. Assuming that
$\Gg_0\ll\GD_0$, we can replace $f(x)$ with a delta function
normalized to the same area:
\begin{equation}\label{Eq:DeltaFunc}
    f(x)
    \simeq\frac{\Gp\Gg_0}{2\GD_0\sqrt{1-D_0^2}}\,\Gd\!\prn{D_0-\cos x}.
\end{equation}
Substituting this expression into Eq.\ (\ref{Eq:anExact}), we
obtain
\begin{equation}\label{Eq:anApprox}
    a_n
    \simeq\frac{\Gg_0}{2\GD_0}
        \frac{\cos\prn{n\arccos{D_0}}}{\sqrt{1-D_0^2}}.
\end{equation}
This approximation breaks down for $\abs{D_0}$ within
$\sim$$\Gg_0/\GD_0$ of unity. However, as we see below, we are
interested in integrals of $a_n$ over effective detuning, which
can be well approximated using the expression (\ref{Eq:anApprox}).
We are also limited by this approximation to harmonics
$n\ll\GD_0/\Gg_0$, since the factor $\cos nx$ is assumed to not
vary rapidly over the optical resonance. Thus, from Eq.\
(\ref{Eq:bnDef}), $b_n$ can be approximated by
\begin{equation}\label{Eq:bnApprox}
\begin{split}
    b_n
    &{}\simeq\frac{\Gg_0}{4\GD_0}
        \frac{\cos\prn{n\arccos{D_0}}}{\sqrt{1-D_0^2}}\\
    &{}\simeq\frac{1}{2}\,a_n.
\end{split}
\end{equation}
Thus we see that $I_2\simeq I_1/2$ and the terms of Eq.\
(\ref{r21Eq}) proportional to $\GO_L$ cancel. Substituting Eq.\
(\ref{Eq:I1Expr}) into Eq.\ (\ref{r21Eq}), we obtain
\begin{equation}\label{Eq:r21}
    \Gr_{21}\simeq
        \frac{4\GO^2N}{3\Gg_0}\!
        \sum_{n=-\infty}^{\infty}
        \frac{a_n\,e^{in\GO_mt}}{\Gg+i\prn{2\GO_L+n\GO_m}}\,.
\end{equation}

The result (\ref{Eq:r21}) applies to atoms that are at rest. We
now modify this result to describe an atomic ensemble with a
Maxwellian velocity distribution leading to a Doppler width
$\GG_{\!D}$ of the transition. For an atomic velocity group with
component of velocity $v$ along the light propagation direction,
the light frequency is shifted according to
$\Go(v)=\Go\prn{1-v/c}=\Go-kv$ where $k$ is the light-field wave
number. Writing the dimensionless Doppler-shift parameter
$D_v=-kv/\GD_0$, the atomic density $N$ for this velocity group
becomes
\begin{equation}\label{Eq:dNv}
    dN(v)
    =\frac{\GD_0}{\GG_{\!D}\sqrt{\Gp}}\,N_0\,
        e^{-\prn{D_v\GD_0/\GG_{\!D}}^2}\,dD_v,
\end{equation}
where $N_0$ is the total atomic density, and the average detuning
parameter $D_0$ becomes $D_0(v)=D_0+D_v$. Defining the
velocity-dependent coefficient $a_n\ms{-1}(v)$ by
\begin{equation}\label{Eq:bnv}
\begin{split}
    a_n\ms{-1}(v)\,dD_v
    &=\frac{\GG_{\!D}}{\Gg_0}\frac{dN(v)}{N_0}\,a_n\\
    &{}\simeq
        \frac{\cos\sbr{n\arccos\prn{D_0+D_v}}}{2\sqrt{\Gp}\sqrt{1-\prn{D_0+D_v}^2}}\,
        e^{-\prn{D_v\GD_0/\GG_{\!D}}^2}dD_v,
\end{split}
\end{equation}
the velocity-dependent ground-state coherence $\Gr_{21}\ms{-1}(v)$
is given by
\begin{equation}\label{Eq:r21v}
\begin{split}
    \Gr_{21}\ms{-1}(v)\,dD_v
    &{}\simeq\frac{4\GO^2\,d\ms{-1}N\ms{-1}(v)}{3\Gg_0}\!
        \sum_{n=-\infty}^{\infty}
        \frac{a_n\,e^{in\GO_mt}}{\Gg+i\prn{2\GO_L+n\GO_m}}\\
    &{}\simeq
        \frac{4\GO^2N_0}{3\GG_{\!D}}\!
        \sum_{n=-\infty}^{\infty}
        \frac{a_n\ms{-1}(v)\,e^{in\GO_mt}}{\Gg+i\prn{2\GO_L+n\GO_m}}\,dD_v.
\end{split}
\end{equation}
In a situation in which atomic collisions are important, such as
in a vapor cell with a buffer gas or an antirelaxation coating,
this result must be further modified to take into account
collisionally induced velocity mixing. For atoms contained in an
antirelaxation-coated vapor cell, we assume that each velocity
group interacts separately with the excitation light, but after
pumping all groups are completely mixed. This model applies when
light power is low enough so that optical pumping averaged over
the atomic velocity distribution and the cell volume is
unsaturated. The ground-state coherence of each velocity group
becomes the velocity-averaged quantity $\bar{\Gr}_{21}\ms{-1}(v)$,
given by the normalized velocity average of Eq.\ (\ref{Eq:r21v}):
\begin{equation}
\begin{split}
    \bar{\Gr}_{21}\ms{-1}(v)\,dD_v
        &{}\simeq\frac{dN(v)}{N_0}
            \int_{-\infty}^{\infty}
                \Gr_{21}\ms{-1}(v)\,
            dD_v\\
        &{}=\frac{4\GO^2\,d\ms{-1}N\ms{-1}(v)}{3\GG_{\!D}}
            \sum_{n=-\infty}^{\infty}
            \frac{\bar{a}_n\,e^{in\GO_mt}}{\Gg+i\prn{2\GO_L+n\GO_m}}\,,
\end{split}
\end{equation}
where the averaged coefficient $\bar{a}_n$ is given by
\begin{equation}\label{Eq:bnAvg}
\begin{split}
    \bar{a}_n
    &=\int_{-\infty}^{\infty}a_n\ms{-1}(v)\,dD_v\\
    &{}\simeq\int_{-\infty}^{\infty}
            \frac{\cos\sbr{n\arccos\prn{D_0+D_v}}}{2\sqrt{\Gp}\sqrt{1-\prn{D_0+D_v}^2}}\,
            e^{-\prn{D_v\GD_0/\GG_{\!D}}^2}\,dD_v.
\end{split}
\end{equation}
Below, we will need the real and imaginary parts of
$\bar{\Gr}_{21}$, given by
\begin{equation}\label{ReImr21}
\begin{split}
    \re{}&\bar{\Gr}_{21}\ms{-1}(v)\,dD_v
    \simeq\frac{4\GO^2\,d\ms{-1}N\ms{-1}(v)}{3\GG_{\!D}}\\
        &{}\times\sum_{n=-\infty}^{\infty}
        \frac{
            \bar{a}_n
            \sbr{
                \Gg\cos n\GO_m t
                +\prn{2\GO_L+n\GO_m}\sin n\GO_m t
            }
        }{\Gg^2+\prn{2\GO_L+n\GO_m}^2}\,
    ,\\
    \im{}&\bar{\Gr}_{21}\ms{-1}(v)\,dD_v
    \simeq\frac{4\GO^2\,d\ms{-1}N\ms{-1}(v)}{3\GG_{\!D}}\\
        &{}\times\sum_{n=-\infty}^{\infty}
        \frac{
            \bar{a}_n
            \sbr{
                \Gg\sin n\GO_m t
                -\prn{2\GO_L+n\GO_m}\cos n\GO_m t
            }
        }{\Gg^2+\prn{2\GO_L+n\GO_m}^2}\,.
\end{split}
\end{equation}

\subsection{Optical properties of the medium}

We now derive the formula for the optical rotation in terms of the
polarization of the medium $\mb{P}=\tr\rho\mb{d}$. The electric
field of coherent light of arbitrary polarization can be described
by \cite{Huard}
\begin{equation}\label{lightfield}
\begin{split}
    \mb{E}
    &{}=\frac{1}{2}
        \sbrk{
            E_0e^{i\Gf}
            \prn{
                \cos\Gv\cos\Ge-i\sin\Gv\sin\Ge}e^{i\prn{\Go t-kz}
            }+c.c.
        }\hat{\mb{e}}_x\\
    &{}+\frac{1}{2}
        \sbrk{
            E_0e^{i\Gf}
            \prn{
                \sin\Gv\cos\Ge+i\cos\Gv\sin\Ge}e^{i\prn{\Go t-kz}
            }+c.c.
        }\hat{\mb{e}}_y,
\end{split}
\end{equation}
where $k=\Go/c$ is the vacuum wave number, $\Gf$ is the overall
phase, $\Gv$ is the polarization angle, and $\Ge$ is the
ellipticity.

Substituting Eq.\ (\ref{lightfield}) into the wave equation
\begin{equation}
    \prn{\frac{d^2}{dz^2}-\frac{d^2}{c^2dt^2}}\mb{E}
    =-\frac{4\Gp}{c^2}\frac{d^2}{dt^2}\mb{P},
\end{equation}
and neglecting terms involving second-order derivatives and
products of first-order derivatives (thus assuming that changes in
$\Gv$, $\Ge$, and $\Gf$ and fractional changes in $E_0$ are
small), gives the rotation, phase shift, absorption, and change of
ellipticity per unit distance:
\begin{equation}
\begin{split}
    \frac{d\Gv}{dz}
    =&-\frac{2\Gp\Go}{E_0c}\sec2\Ge
        \left[
            \cos\Gv\prn{P_1\sin\Ge+P_4\cos\Ge}
        \right.\\
        &\ms{150}\left.
            {}-\sin\Gv\prn{P_2\cos\Ge-P_3\sin\Ge}
        \right],\\
    \frac{d\Gf}{dz}
    =&-\frac{2\Gp\Go}{E_0c}\sec2\Ge
        \left[
            \cos\Gv\prn{P_1\cos\Ge+P_4\sin\Ge}
        \right.\\
        &\ms{150}\left.
            {}-\sin\Gv\prn{P_2\sin\Ge-P_3\cos\Ge}
        \right],\\
    \frac{dE_0}{dz}
    =&\frac{2\Gp\Go}{c}
        \left[
            \sin\Gv\prn{P_1\sin\Ge-P_4\cos\Ge}
        \right.\\
        &\ms{150}\left.
            {}-\cos\Gv\prn{P_2\cos\Ge+P_3\sin\Ge}
        \right],\\
    \frac{d\Ge}{dz}
    =&\frac{2\Gp\Go}{E_0c}
        \left[
            \sin\Gv\prn{P_1\cos\Ge+P_4\sin\Ge}
        \right.\\
        &\ms{150}\left.
            {}+\cos\Gv\prn{P_2\sin\Ge-P_3\cos\Ge}
        \right],
\end{split}
\end{equation}
where the components $P_{1,2,3,4}$ of the polarization are defined
by
\begin{equation}
\begin{split}
    \mb{P}
    &{}=\frac{1}{2}
        \sbr{
            \prn{P_1-iP_2}e^{i\prn{\Go t-kz}}+c.c.
        }\hat{\mb{e}}_x\\
    &{}+\frac{1}{2}
        \sbr{
            \prn{P_3-iP_4}e^{i\prn{\Go t-kz}}+c.c.
        }\hat{\mb{e}}_y.
\end{split}
\end{equation}
For initial values of $\Gv=\Ge=0$, the rotation per unit length is
given by
\begin{equation}\label{Eq:OptRot0}
    \frac{d\Gv}{d\ell}
    =-\frac{2\Gp\Go P_4}{cE_0}\,.
\end{equation}

\subsection{Calculation of the signal}
\label{Sec:Signal}

We now evaluate $\mb{P}=\tr\Gr\mb{d}$ and substitute into Eq.\
(\ref{Eq:OptRot0}) to find the optical rotation in terms of the
ground-state atomic coherence derived above. Taking into account
that in the nonrotating frame the optical atomic coherences
oscillate at the light frequency $\Go$, we find for the
polarization components
\begin{equation}
\begin{split}
    P_1&=\sqrt{\frac{2}{3}}\rme{\xi_gJ_g}{d}{\xi_eJ_e}\re\prn{\Gr_{31}-\Gr_{23}},\\
    P_2&=-\sqrt{\frac{2}{3}}\rme{\xi_gJ_g}{d}{\xi_eJ_e}\im\prn{\Gr_{31}+\Gr_{23}},\\
    P_3&=-\sqrt{\frac{2}{3}}\rme{\xi_gJ_g}{d}{\xi_eJ_e}\im\prn{\Gr_{31}-\Gr_{23}},\\
    P_4&=-\sqrt{\frac{2}{3}}\rme{\xi_gJ_g}{d}{\xi_eJ_e}\re\prn{\Gr_{31}+\Gr_{23}},
\end{split}
\end{equation}
so the optical rotation angle per unit length is given by
\begin{equation}\label{OptRot}
\begin{split}
    \frac{d\Gv}{d\ell}
    &=\frac{\Gp\Go\rme{\xi_gJ_g}{d}{\xi_eJ_e}^2}{3\GO c}\re\prn{\Gr_{31}+\Gr_{32}}\\
    &=\frac{\Gg_0\Gl^2}{16\Gp\GO}\re\prn{\Gr_{31}+\Gr_{32}},
\end{split}
\end{equation}
where $\Gl$ is the transition wavelength. Here we have used the
fact that for a closed $J\ra J'$ transition \cite{Sob92},
\begin{equation}
    \Gg_0=\frac{4\ms{1}\Go_0^3}{3\ms{1}c^3}\frac{1}{2J'+1}\rme{\xi\!J}{d}{\xi'\ms{-4}J'}^2,
\end{equation}
and that $\Go\simeq\Go_0$.

Substituting in the expressions (\ref{OptCoherence}), and assuming
$\Gr_{11}\simeq\Gr_{22}\simeq N/3$ and $\GO_L\ll\Gg_0$, Eq.\
(\ref{OptRot}) can be written in terms of the ground-state
coherence as
\begin{widetext}
\begin{equation}\label{OptRot1}
    \frac{d\Gv}{d\ell}
    =\frac{\Gg_0\Gl^2}{8\Gp}
        \prn{
            \frac{\GO_L\prn{N/3-\re\Gr_{21}}+\prn{\Gg_0/2}\im\Gr_{21}}{\Gg_0^2/4+\GD\ms{-1}^2}
            -\frac{2\GO_L\GD\ms{-1}^2\prn{N/3-\re\Gr_{21}}}{\prn{\Gg_0^2/4+\GD\ms{-1}^2}^2}
        },
\end{equation}
or, for the case of complete velocity mixing:
\begin{equation}\label{OptRot1Avg}
    \frac{d\Gv(v)}{d\ell}\,dD_v
    =\frac{\Gg_0\Gl^2}{8\Gp}
        \prn{
            \frac{\GO_L\sbr{dN(v)/3-\re\bar{\Gr}_{21}\ms{-1}(v)\,dD_v}+\prn{\Gg_0/2}\im\bar{\Gr}_{21}\ms{-1}(v)\,dD_v}{\Gg_0^2/4+\GD\ms{-1}^2(v)}
            -\frac{2\GO_L\GD\ms{-1}^2(v)\sbr{dN(v)/3-\re\bar{\Gr}_{21}\ms{-1}(v)\,dD_v}}{\sbr{\Gg_0^2/4+\GD\ms{-1}^2(v)}^2}
        },
\end{equation}
\end{widetext}
where the velocity-dependent effective detuning $\GD(v)$ is given,
as before, by
\begin{equation}
    \GD(v)=\GD_0\sbr{D_0(v)-\cos\GO_m\Gt}.
\end{equation}

The in-phase and quadrature signals (see Sec.\
\ref{Sec:DataAndTheory}) per unit length of the medium, measured
for a time $T$ at the $j$-th harmonic of the modulation frequency,
are given by the time averages
\begin{equation}\label{SigEq}
\begin{split}
    \frac{d\Gv^\text{in}_{\!j}(v)}{d\ell}\,dD_v
    &{}=\frac{dD_v}{T}\int_0^T\frac{d\Gv(v)}{d\ell}\cos\prn{j\,\GO_mt}dt,\\
    \frac{d\Gv^\text{out}_{\!j}(v)}{d\ell}\,dD_v
    &{}=\frac{dD_v}{T}\int_0^T\frac{d\Gv(v)}{d\ell}\sin\prn{j\,\GO_mt}dt.
\end{split}
\end{equation}
We substitute the formulas for the real and imaginary parts of the
ground-state coherence [Eq.\ (\ref{ReImr21})] into the formula for
the optical rotation [Eq.\ (\ref{OptRot1Avg})], and the resulting
expression into Eq.\ (\ref{SigEq}). After evaluating the time
integrals (see Appendix \ref{Sec:TimeInt}), we find that the
signals due to each velocity group are given by
\begin{equation}\label{Eq:sigv1}
\begin{split}
    \frac{d\Gv^\text{in}_{\!j}(v)}{d\ell}\,dD_v
    \simeq{}&-\frac{\Gl^2\GO^2}{6\Gp\GG_{\!D}}\,dN(v)\\
        &{}\times\sum_{n=-\infty}^{\infty}\ms{-6}
            \frac{
                \prn{2\GO_L+n\GO_m}
                \bar{a}_n\prn{a_{n+j}+a_{n-j}}
            }
            {\Gg^2+\prn{2\GO_L+n\GO_m}^2},\\
    \frac{d\Gv^\text{out}_{\!j}(v)}{d\ell}\,dD_v
    \simeq{}&-\frac{\Gl^2\GO^2}{6\Gp\GG_{\!D}}\,dN(v)\\
        &{}\times\sum_{n=-\infty}^{\infty}\ms{-6}
            \frac{
                \Gg\,
                \bar{a}_n\prn{a_{n+j}-a_{n-j}}
            }
            {\Gg^2+\prn{2\GO_L+n\GO_m}^2}\,.
\end{split}
\end{equation}
Using the definitions of $dN(v)$ and $a_n\ms{-1}(v)$ [Eqs.\
(\ref{Eq:dNv},\ref{Eq:bnv})] we can rewrite Eq.\ (\ref{Eq:sigv1})
as
\begin{equation}\label{Eq:sigv2}
\begin{split}
    &\frac{d\Gv^\text{in}_{\!j}(v)}{d\ell}\,dD_v\\
    &\ms{10}{}\simeq\Gh\!\!
        \sum_{n=-\infty}^{\infty}\ms{-6}
            \frac{
                \Gg\prn{2\GO_L+n\GO_m}
                \bar{a}_n\sbr{a_{n+j}\ms{-1}(v)+a_{n-j}\ms{-1}(v)}
            }
            {\Gg^2+\prn{2\GO_L+n\GO_m}^2}\,dD_v,\\
    &\frac{d\Gv^\text{out}_{\!j}(v)}{d\ell}\,dD_v\\
    &\ms{10}{}\simeq\Gh\!\!
        \sum_{n=-\infty}^{\infty}\ms{-6}
            \frac{
                \Gg^2\,
                \bar{a}_n\sbr{a_{n+j}\ms{-1}(v)-a_{n-j}\ms{-1}(v)}
            }
            {\Gg^2+\prn{2\GO_L+n\GO_m}^2}\,\,dD_v,
\end{split}
\end{equation}
where the signal amplitude factor $\Gh$ is defined by
\begin{equation}
    \Gh=-\frac{1}{6\Gp}\frac{\GO^2\Gg_0}{\GG_{\!D}^2\Gg}\,\Gl^2\!N_0.
\end{equation}
The total signal, given by the integral over all velocity groups,
is found by replacing $a_n\ms{-1}(v)$ with $\bar{a}_n$:
\begin{equation}\label{Eq:sigTot}
\begin{split}
    \rb{\frac{d\Gv^\text{in}_{\!j}}{d\ell}}_\text{total}
    &=\int_{-\infty}^{\infty}\frac{d\Gv_\text{in}(v)}{d\ell}\,dD_v\\
    &\simeq\Gh\!\!
        \sum_{n=-\infty}^{\infty}\ms{-6}
            \frac{
                \Gg\prn{2\GO_L+n\GO_m}
                \bar{a}_n\prn{\bar{a}_{n+j}+\bar{a}_{n-j}}
            }
            {\Gg^2+\prn{2\GO_L+n\GO_m}^2},\\
    \rb{\frac{d\Gv^\text{out}_{\!j}}{d\ell}}_\text{total}
    &=\int_{-\infty}^{\infty}\frac{d\Gv_\text{out}(v)}{d\ell}\,dD_v\\
    &\simeq\Gh\!\!
        \sum_{n=-\infty}^{\infty}\ms{-6}
            \frac{
                \Gg^2\,
                \bar{a}_n\prn{\bar{a}_{n+j}-\bar{a}_{n-j}}
            }
            {\Gg^2+\prn{2\GO_L+n\GO_m}^2}\,.
\end{split}
\end{equation}
Each term of the sums corresponds to a resonance at
$\GO_L/\GO_m=-n/2$ (Figs.\
\ref{Fig:1HDataTheoryPlot},\ref{Fig:2HDataTheoryPlot}). Near each
resonance the in-phase signal is dispersive in shape, whereas the
quadrature signal is a Lorentzian. When plotted as a function of
the Larmor frequency normalized to the modulation frequency,
$\GO_L/\GO_m$, the widths of the resonances are determined by the
normalized ground-state relaxation rate $\Gg/\GO_m$. The relative
amplitudes of the resonances are determined by the ratio of the
modulation depth to the Doppler width, $\GD_0/\GG_{\!D}$, and the
normalized average detuning $\GD_l/\GD_0$.

\section{Conclusion}
\label{Sec:Conclusion}

We have presented a theory of nonlinear magneto-optical rotation
with low-power frequency-modulated light for a
low-angular-momentum system. The magnetic-field dependence
predicted by this theory is in qualitative agreement with
experimental data taken on the Rb $D1$ line. Directions for future
work include a more complete theory describing
higher-angular-momentum systems, including systems with hyperfine
structure, and higher light powers. A possible complication to the
FM NMOR technique in systems with hyperfine structure is the
nonlinear Zeeman effect present at higher magnetic fields, so a
theoretical description of this effect would also be helpful. FM
NMOR has been shown to be a useful technique for the selective
study of higher-order polarization moments, which give rise to
distinct resonances at different values of the magnetic field than
the quadrupole resonances studied here \cite{Yas2003Select} (see
also Ref.\ \cite{Auz84}). Higher-order moments are of interest in
part because signals due to the highest-order moments possible in
a given system would be free of the complications due to the
nonlinear Zeeman effect. To describe these moments, a calculation
along the same lines as the one presented here but carried out to
higher order and involving more atomic sublevels would be
necessary.

\acknowledgments

We thank M. Auzinsh, W. Gawlik, and A. Lezama for helpful
discussions. This work has been supported by the Office of Naval
Research (grant N00014-97-1-0214); by a US-Armenian bilateral
Grant CRDF AP2-3213/NFSAT PH 071-02; by NSF; by the Director,
Office of Science, Nuclear Science Division, of the U.S.
Department of Energy under contract DE-AC03-76SF00098; and by a
CalSpace Minigrant. D.B. also acknowledges the support of the
Miller Institute for Basic Research in Science.

\appendix

\section{Evaluation of the time integrals}
\label{Sec:TimeInt}

In evaluating Eq.\ (\ref{SigEq}), several time integrals appear:
\begin{equation*}
    I_3
    =\frac{1}{\GO_mT}
        \int_0^{\GO_mT}\frac{\cos jx}{\Gg_0^2/4+\GD_0^2\sbr{D_0(v)-\cos x}^2}\,dx,
\end{equation*}
\begin{equation*}
    I_4
    =\frac{1}{\GO_mT}
        \int_0^{\GO_mT}\frac{\sin jx}{\Gg_0^2/4+\GD_0^2\sbr{D_0(v)-\cos x}^2}\,dx,
\end{equation*}
\begin{equation*}
    I_5
    =\frac{1}{\GO_mT}
        \int_0^{\GO_mT}\frac{\cos jx\sin nx}{\Gg_0^2/4+\GD_0^2\sbr{D_0(v)-\cos x}^2}\,dx,
\end{equation*}
\begin{equation*}
    I_6
    =\frac{1}{\GO_mT}
        \int_0^{\GO_mT}\frac{\sin jx\sin nx}{\Gg_0^2/4+\GD_0^2\sbr{D_0(v)-\cos x}^2}\,dx,
\end{equation*}
\begin{equation*}
    I_7
    =\frac{1}{\GO_mT}
        \int_0^{\GO_mT}\frac{\cos jx\cos nx}{\Gg_0^2/4+\GD_0^2\sbr{D_0(v)-\cos x}^2}\,dx,
\end{equation*}
\begin{equation*}
    I_8
    =\frac{1}{\GO_mT}
        \int_0^{\GO_mT}\frac{\sin jx\cos nx}{\Gg_0^2/4+\GD_0^2\sbr{D_0(v)-\cos x}^2}\,dx,
\end{equation*}
as well as the related integrals
\begin{equation*}
\begin{split}
    I_9
    &=\frac{1}{\GO_mT}
        \int_0^{\GO_mT}
            \frac{\GD_0^2\sbr{D_0(v)-\cos x}^2\cos jx}{\cbr{\Gg_0^2/4+\GD_0^2\sbr{D_0(v)-\cos x}^2}^2}\,
        dx\\
    &=-\GD_0^2\frac{\partial I_3}{\partial\GD_0^2},
\end{split}
\end{equation*}
\begin{equation*}
\begin{split}
    I_{10}
    &=\frac{1}{\GO_mT}
        \int_0^{\GO_mT}
            \frac{\GD_0^2\sbr{D_0(v)-\cos x}^2\sin jx}{\cbr{\Gg_0^2/4+\GD_0^2\sbr{D_0(v)-\cos x}^2}^2}\,
        dx\\
    &=-\GD_0^2\frac{\partial I_4}{\partial\GD_0^2},
\end{split}
\end{equation*}
\begin{equation*}
\begin{split}
    I_{11}
    &=\frac{1}{\GO_mT}
        \int_0^{\GO_mT}
            \frac{\GD_0^2\sbr{D_0(v)-\cos x}^2\cos jx\sin nx}{\cbr{\Gg_0^2/4+\GD_0^2\sbr{D_0(v)-\cos x}^2}^2}\,
        dx\\
    &=-\GD_0^2\frac{\partial I_5}{\partial\GD_0^2},
\end{split}
\end{equation*}
\begin{equation*}
\begin{split}
    I_{12}
    &=\frac{1}{\GO_mT}
        \int_0^{\GO_mT}
            \frac{\GD_0^2\sbr{D_0(v)-\cos x}^2\sin jx\sin nx}{\cbr{\Gg_0^2/4+\GD_0^2\sbr{D_0(v)-\cos x}^2}^2}\,
        dx\\
    &=-\GD_0^2\frac{\partial I_6}{\partial\GD_0^2},
\end{split}
\end{equation*}
\begin{equation*}
\begin{split}
    I_{13}
    &=\frac{1}{\GO_mT}
        \int_0^{\GO_mT}
            \frac{\GD_0^2\sbr{D_0(v)-\cos x}^2\cos jx\cos nx}{\cbr{\Gg_0^2/4+\GD_0^2\sbr{D_0(v)-\cos x}^2}^2}\,
        dx\\
    &=-\GD_0^2\frac{\partial I_7}{\partial\GD_0^2},
\end{split}
\end{equation*}
\begin{equation*}
\begin{split}
    I_{14}
    &=\frac{1}{\GO_mT}
        \int_0^{\GO_mT}
            \frac{\GD_0^2\sbr{D_0(v)-\cos x}^2\sin jx\cos nx}{\cbr{\Gg_0^2/4+\GD_0^2\sbr{D_0(v)-\cos x}^2}^2}\,
        dx\\
    &=-\GD_0^2\frac{\partial I_8}{\partial\GD_0^2}.
\end{split}
\end{equation*}
If $T$ is many modulation periods, the above integrals can be
approximated by averages over one period. Thus, we can change the
limits of the integrals to $\prn{-\Gp,\Gp}$, and set the
normalizing factor $\prn{\GO_mT}^{-1}$ to $\prn{2\Gp}^{-1}$. Using
the trigonometric substitutions
\begin{equation}
\begin{split}
    \cos jx\cos nx
    &{}=\frac{1}{2}\cbr{\cos\sbr{\prn{n-j}x}+\cos\sbr{\prn{n+j}x}},\\
    \cos jx\sin nx
    &{}=\frac{1}{2}\cbr{\sin\sbr{\prn{n-j}x}+\sin\sbr{\prn{n+j}x}},\\
    \sin jx\sin nx
    &{}=\frac{1}{2}\cbr{\cos\sbr{\prn{n-j}x}-\cos\sbr{\prn{n+j}x}},\\
    \sin jx\cos nx
    &{}=-\frac{1}{2}\cbr{\sin\sbr{\prn{n-j}x}+\sin\sbr{\prn{n+j}x}},
\end{split}
\end{equation}
we can rewrite the above integrals in terms of the $a_n$ and $b_n$
coefficients.
\begin{equation}
\begin{split}
    I_3&{}\simeq\prn{\Gg_0^2/4}^{-1}a_j,\\
    I_4&{}\simeq0,\\
    I_5&{}\simeq0,\\
    I_6&{}\simeq\prn{\Gg_0^2/4}^{-1}\frac{\prn{a_{n-j}-a_{n+j}}}{2},\\
    I_7&{}\simeq\prn{\Gg_0^2/4}^{-1}\frac{\prn{a_{n-j}+a_{n+j}}}{2},\\
    I_8&{}\simeq0,\\
    I_9&{}\simeq\prn{\Gg_0^2/4}^{-1}b_j,\\
    I_{10}&{}\simeq0,\\
    I_{11}&{}\simeq0,\\
    I_{12}&{}\simeq\prn{\Gg_0^2/4}^{-1}\frac{\prn{b_{n-j}-b_{n+j}}}{2},\\
    I_{13}&{}\simeq\prn{\Gg_0^2/4}^{-1}\frac{\prn{b_{n-j}+b_{n+j}}}{2},\\
    I_{14}&{}\simeq0.
\end{split}
\end{equation}
As in the evaluation of Eq.\ (\ref{r21Eq}), use of the approximate
expression $b_n\simeq a_n/2$ [Eq.\ (\ref{Eq:bnApprox})] results in
the cancellation of some terms proportional to $\GO_L$, producing
the relatively simple form of Eq.\ (\ref{Eq:sigv1}).

\bigskip
\bigskip
\bigskip
\bigskip
\bigskip
\bigskip
\bigskip
\bigskip

~

\bibliography{NMObibl}

\end{document}